\documentclass[12pt,a4paper]{paper}
 \usepackage[margin=2cm]{geometry}
\usepackage{graphicx} 
\usepackage{paralist}
\usepackage{amsmath, amsfonts, amssymb}
\usepackage{amsbsy}
\usepackage{epsfig}
\usepackage{setspace}
\usepackage{color}
\usepackage{subfig}
\usepackage{paralist}
\usepackage{bm}
\usepackage{cancel}
\usepackage{appendix}

\newcommand{\p }{ \mathbb{P} }
\newcommand{\me}{\mathrm{e}}

\newtheorem{definition}{Definition}[section]

\title{Asymptotic frequency of shapes in supercritical branching trees}
\author{Giacomo Plazzotta\thanks{Imperial College London. Corresponding author: giacomo.plazzotta11@imperial.ac.uk} \& Caroline Colijn\thanks{Imperial College London}}

\begin{document}
\maketitle
\abstract{The shapes of branching trees have been linked to disease transmission patterns. In this paper we use the general Crump-Mode-Jagers branching process to model an outbreak of an infectious disease under mild assumptions. Introducing a new class of characteristic functions, we are able to derive a formula for the limit of the frequency of the occurrences of a given shape in a general tree. The computational challenges concerning the evaluation of this formula are in part overcome using the Jumping Chronological Contour Process. We apply the formula to derive the limit of the frequency of cherries, pitchforks and double cherries in the constant rate birth-death model, and the frequency of cherries under a non-constant death rate.}

\section{Introduction}
Branching processes are widely studied and used to model many biological growth phenomena. Although their first and most direct application has been in the study of evolution and extinction \cite{jagers1975branching}, they have also been employed to model infectious disease epidemics \cite{holmes1995revealing, wilson2005germs}. In this context each branching event represents an infection at which a new infectious individual enters the model. The individual's history corresponds to a path in the tree, beginning at the time of infection and ending at a tip which represents the individual's death or recovery.

In recent years, improvements in sequencing technologies have made it possible to  detect micro-evolutionary events in pathogens. Pathogen sequence data can be used to infer  branching trees which in turn can inform our understanding of the disease's transmission dynamics \cite{didelot2014bayesian, kato2013use, ypma2013relating}. The inference of trees from sequence data becomes more challenging as more and more isolates (tips in the tree) are sequenced, presenting significant challenges for the field. Inference relies on tree likelihoods, which are derived from branching processes  \cite{drummond2007beast, stadler2009incomplete, lambert2014phylogenetic}. The shapes of phylogenetic trees have also been linked to disease transmission patterns \cite{frost2013modelling, colijn2014phylogenetic, poon2013mapping}.

The shape of a tree can be defined, informally, as the tree without considering the associated branch lengths \cite{gernhard2008stochastic}. The shape distribution for the Yule model was studied originally in order to estimate the branching points of a branching diffusion process \cite{edwards1970estimation}. Similarly, numerous studies can be found on tree likelihoods and inference, though most exploit the timing of branching events in trees rather than focusing on tree shapes. Moreover, the Yule tree is the most well-studied model, for which there are results describing internal structure and shape distribution \cite{cavalli1967phylogenetic, harding1971probabilities, page1991random, brown1994probabilities}. Hence the shape distribution, at least for the homogeneous (Yule) model, can be considered as resolved. For non-homogeneous models however, few results are available in the literature.

The frequency of a shape in a tree is the ratio between the number of occurrences of that particular shape and the number of tips in the tree.  Some results regarding shape frequency are available for simple tree models \cite{mckenzie2000distributions, rosenberg2006mean, chang2010limit}. In more general settings, the limit of the frequency of a shape will depend on the process defining the tree. Therefore, shape frequencies could provide a tool to estimate the governing parameters of a branching process, using trees derived from empirical data. Given the intractable number of tree shapes with $n$ tips \cite{harding1971probabilities} ($(2n-3)\cdot(2n-1)...3\cdot1$), computational approaches to finding the frequencies of small shape patterns in large trees can have limited success. This motivates an analytical derivation of shape frequencies.

In this work we use the Crump-Mode-Jagers branching process (CMJ process) \cite{jagers1975branching} to obtain the asymptotic frequencies of tree shapes. The CMJ process is a general model with very mild assumptions; it  includes as special cases the Yule process and the homogeneous (constant-rate birth-death) process. We focus on supercritical trees, whose Malthusian parameter is $\geq 1$, because they have a positive probability of never reaching extinction. Using novel characteristic functions and previous convergence properties of the CMJ \cite{nerman1981convergence}, we derive a general formula for the asymptotic frequency of potentially any shape in any tree. The evaluation of the formula presents computational challenges, which we  overcome in part by applying the Jumping Chronological Contour Process \cite{lambert2008population, lambert2010contour}. We evaluate the expression for the asymptotic frequency of some simple shapes in the homogeneous tree and a non-homogeneous model.






\section{Background}\label{sec:background}
The theory of general Crump-Mode-Jagers (CMJ) branching processes provides an ideal framework for the study of sub-shapes because they can be counted with characteristic functions. We begin with some definitions and the results we have used from the literature on CMJ processes. More details can be found in \cite{jagers1975branching, athreya1972branching}.

\subsubsection*{Notation and definition}
\begin{figure}
	\begin{center}
		\includegraphics[width=0.8\textwidth]{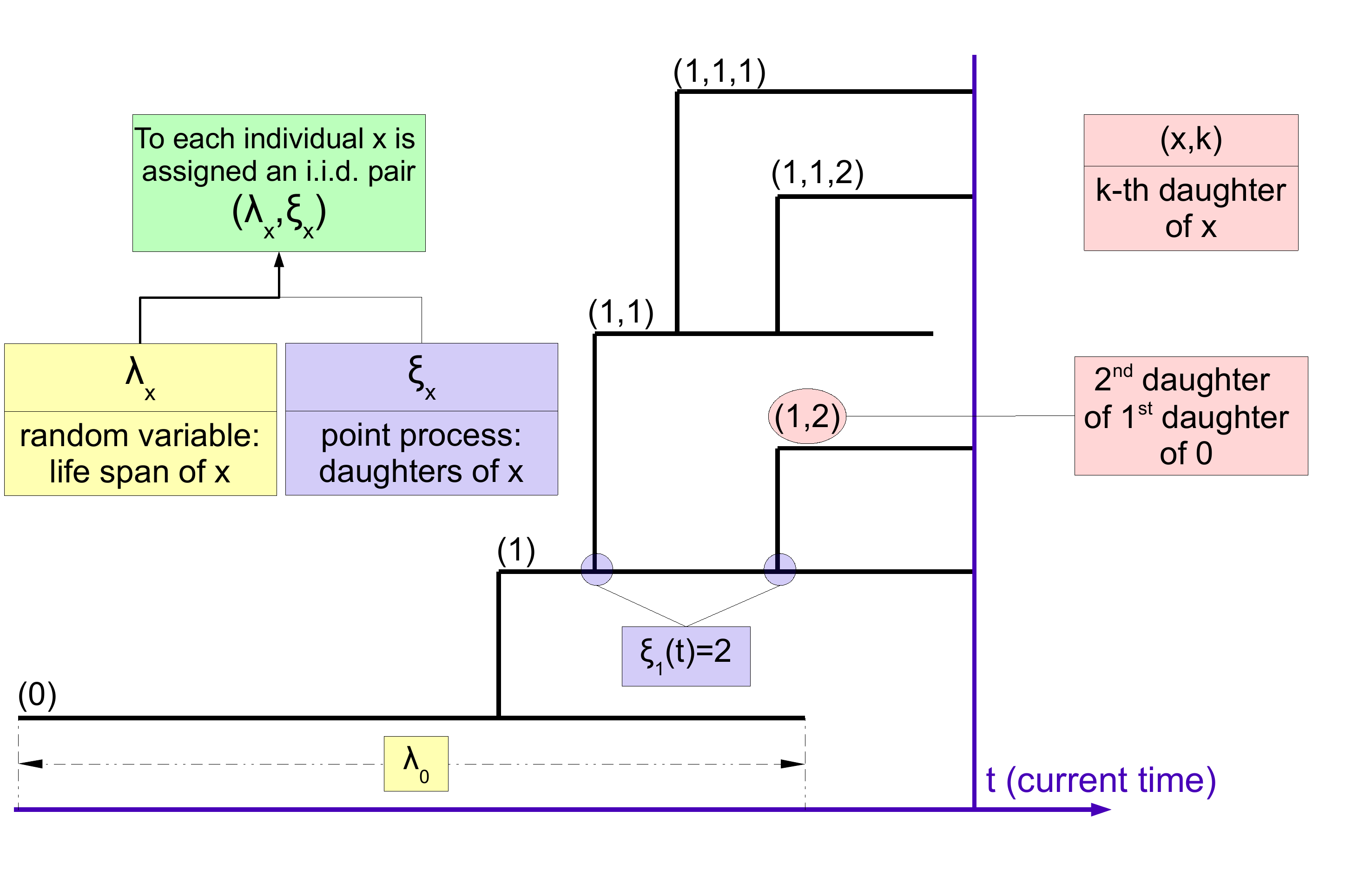}
	\end{center}
	\caption{Explanatory figure that shows the choice of notation in a simple tree generated from a CMJ process}\label{fig:exampletree}
\end{figure}

Following Jagers' setting \cite{jagers1975branching}, each individual of the process is assigned a sequence $x$ in the space $I$ of all the possible sequences of non-negative integers. The sequence $x$ is chosen uniquely in the following way: if the individual with sequence $x$ is the $k$-th daughter of the individual with sequence $y$, then $x=(y,k)$. Therefore, setting the ancestor's sequence as the singleton 0, each individual's  unique sequence keeps track of its predecessors. Since every sequence starts with a 0, for simplicity we will omit this leading 0 from the notation. 

Each individual $x$ is also assigned a random variable $\lambda_x$, the life length of $x$, and a point process $\xi_x$, the reproduction of $x$. Further, it is assumed that the pairs $(\lambda_x, \xi_x)$ are i.i.d. but $\lambda_x$ and $\xi_x$ may depend on each other. For the scope of this paper we need to make two restrictions on the point process $\xi$. First, to have a binary tree, we require that $\xi$ has no multiple points. Second, let $\mu(t)=E[\xi(t)]$: for to have supercriticality of the branching process we require $\lim_{t\to\infty}\mu(t)>1$.

We refer to the constant-rate birth-death process as the homogeneous process. It is a special case of the CMJ process: births happen at a constant rate $\beta$ and deaths at a constant rate $\delta$. In the CMJ setting, this corresponds to the point process $\xi$ being a Poisson process with intensity $\beta$ over the life span $\lambda$, which is exponentially distributed with parameter $\delta$. 

To complete the definition of the branching process, consider the function $z_x(t)$ that indicates whether individual $x$ is alive at time $t$:
$$z_x(t)=\begin{cases} 1 & \text{ if $x$ is alive at time $t$}\\
0 & \text{ otherwise}
\end{cases}.$$
Then the CMJ process $\{Z(t); t\in\mathbb{R}_0^+\}$ is defined as follows:
$$Z(t)=\sum_{x\in I}z_x(t).$$

\subsubsection*{Convergence of CMJ processes}
Because of the self-similar structure of branching processes, renewal theory is often used to analyze them; we refer to \cite{cox1962renewal,jagers1969renewal,jagers1975branching} for detailed applications and proofs. Because we want to analyze the asymptotic behaviour of cherries in supercritical branching processes, it is important to state that such processes converge \cite{jagers1969renewal}, in the sense that:
$$E[\me^{-Mt}Z(t)]\to\frac{\int_0^{\infty}\me^{-M t}E[z_0(t)]\text{d}t}{\int_0^\infty t\me^{-M t}\mu(\text{d}t)},$$
where $\mu(t)=E[\xi(t)]$ and $M$ is the Malthusian parameter of the process, i.e. the positive real number that satisfies $\int_0^\infty \me^{-Mt}\mu(\text{d}t)=1$.

\subsubsection*{The Jumping Chronological Contour Process}
\begin{figure}
	\begin{center}
		\includegraphics[width=0.8\textwidth]{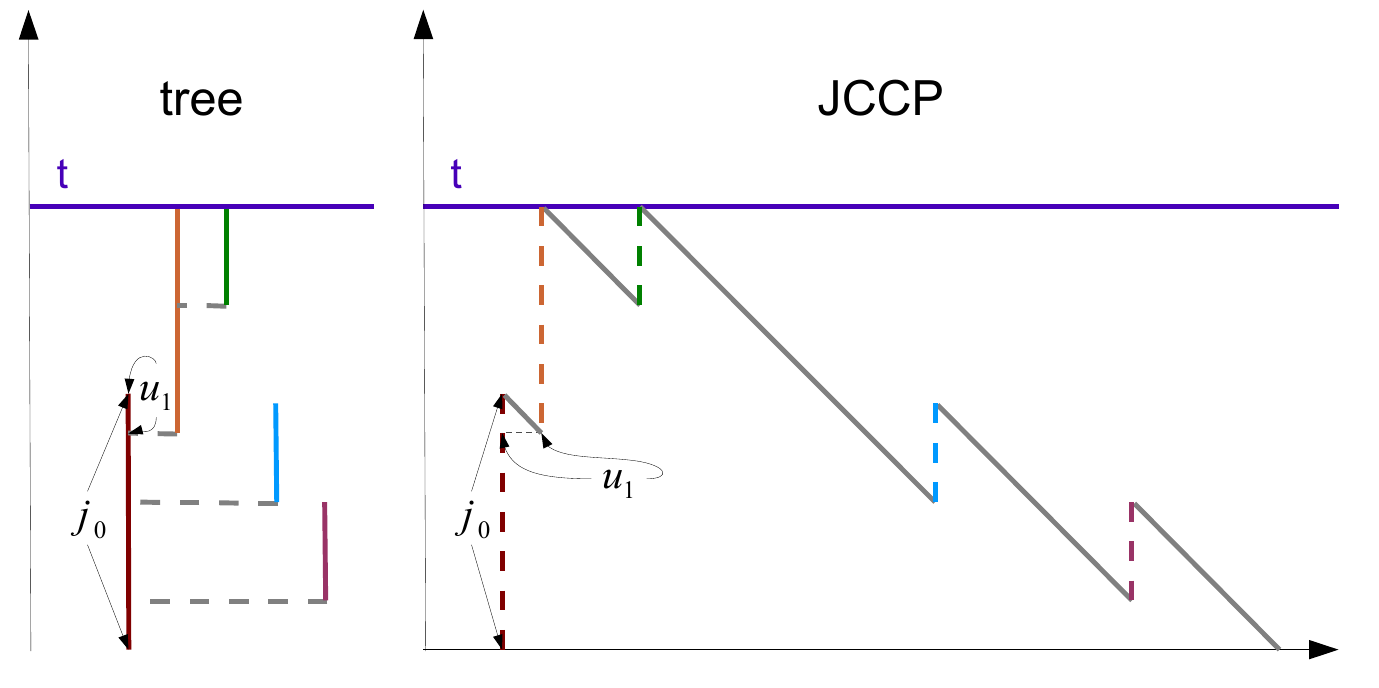}
	\end{center}
	\caption{An example of a Jumping Chronological Contour Process and its relative tree.}\label{fig:examplejccp}
\end{figure}
Contour processes can be interpreted as ``distance to the root'' processes \cite{geiger1995contour}, given a proper distance. We use the Jumping Chronological Contour Process (JCCP) defined in \cite{lambert2008population, lambert2010contour}. To avoid unecessary heavy notation we give an informal description here. Consider a ball that visits every point on the tree starting from the death of the ancestor. The ball proceeds with speed -1 along the lifetime of the individual it is visiting, and when it encounters a birth event (of a descendant of that individual), it jumps to the point representing the death of the newborn. When the ball reaches the birth time of the individual it is visiting, it continues up the tree to visit the mother of that individual (and her descendants, and so on). The process is stopped when the ball reaches the point where the ancestor is born. Figure~\ref{fig:examplejccp}  shows the JCCP and its corresponding tree. The JCCP can be very useful because of the independence of its defining components (jumps and declines) for all trees with constant birth rates. 

Construction of the JCCP can be achieved independently from a previously defined tree through simulation of the jumps $j_i$ and the drops $u_i$ (see Figure~\ref{fig:examplejccp}). When constructing the JCCP two rules need to be applied: reflection and killing upon hitting 0. The first occurs when a jump overshoots the current time $t$ and it is reflected or ``sent back to'' the current time. The second is used to to set an end for the process whenever it hits the $x$-axis. With these two properties, the JCCP has the same law as the tree \cite{lambert2014phylogenetic}.

\section{Characteristic functions for shapes}\label{sec:shapeCharacteristic}
The shape $\mathcal{S}$ of a tree or a subtree can be defined as the tree or the subtree without the associated branch lengths \cite{gernhard2008stochastic}. Common small (low number of tips) shapes are the cherry, the three-tip shape known as a pitchfork, the four-tip symmetric shape called a ``double cherry'', and so on. If the shape $\mathcal{S}$ occurs inside a tree, it is clear that there is a unique individual which is both a tip and an ancestor of $\mathcal{S}$, for each occurrence of $\mathcal{S}$ in the tree. If $x$ is the ancestor of shape $\mathcal{S}$ we say that $x$ \emph{mothers} $\mathcal{S}$. 
\begin{definition}
The characteristic of the shape $\mathcal{S}$ is
\begin{equation}\label{eq:shapeCharacteristic}
	\phi^{\mathcal{S}}_x(t)=\begin{cases}1&\text{ if }x\text{ mothers and is a tip of }\mathcal{S} \\ 0&\text{ otherwise}\end{cases}.
\end{equation}
\end{definition}
The total number of occurrences of shape $\mathcal{S}$  in the tree at time $t$ is $Z^\mathcal{S}(t)=\sum_{x\in I}\phi^\mathcal{S}_x(t)$. For simplicity, if the suffix $x$ is not specified for the characteristic, we imply the ancestor. Note that the characteristic may not be independent between individuals, but we require that it only depends on $x$'s life and on its daughter process. Henceforth we shall assume the following conditions:
\begin{equation}\label{eq:assumption1ForConvergence}
\sum_{k=0}^\infty \sup_{k\leq t\leq t+1}(\me^{-M t} E[\phi(t)])<\infty,
\end{equation}
\begin{equation}\label{eq:assumption2ForConvergence}
E\left[\sup_{s\leq t}\phi(s)\right]<\infty,\text{ for all }t<\infty.
\end{equation}
Statements (\ref{eq:assumption1ForConvergence}) and (\ref{eq:assumption2ForConvergence}) were used in \cite{nerman1981convergence} to prove a number of convergence properties of supercritical trees. In general the number of occurrences of a particular configuration (shape) does not converge in a supercritical tree. However, given two shapes $\mathcal{S}^1$ and $\mathcal{S}^2$ with characteristics $\phi^1$ and $\phi^2$  satisfying equations (\ref{eq:assumption1ForConvergence}) and (\ref{eq:assumption2ForConvergence}), the ratio of the numbers of these two shapes does converge. From \cite{nerman1981convergence} we have: 
\begin{equation}\label{eq:characteristicRatio}
\frac{Z^{\mathcal{S}^1}(t)}{Z^{\mathcal{S}^2}(t)}\to\frac{\int_0^\infty\me^{-Mt}E[\phi^1(t)]\text{d}t}{\int_0^\infty\me^{-Mt}E[\phi^2(t)]\text{d}t},\text{ in probability as }t\to\infty. 
\end{equation}
In equation (\ref{eq:characteristicRatio}) the characteristics $\phi^1$ and $\phi^2$ are associated with the ancestor, i.e. $\phi^{1}:=\phi^{1}_0$ and $\phi^{2}:=\phi^{2}_0$. This is to ensure that the time variable in the characteristic and in the integral are the same, without delays.

The fact that the convergence expressed in equation (\ref{eq:characteristicRatio}) is \emph{in probability} implies that as the tree grows large, the ratio between the occurrences of shapes  $\mathcal{S}^1$ and $\mathcal{S}^2$ tends to its limit (as opposed to convergence in the mean). This holds for any tree from the process. The variance of the shape occurrence ratio calculated in a group of different trees with the same age $t$ tends to zero as $t\to\infty$. This property is very appealing when analyzing large trees, as the difference between the shape occurrence ratio and its limit may be considered statistically insignificant. 

In particular, if we are interested in the frequency of $\mathcal{S}$ (the number of occurrences of shape $\mathcal{S}$ per tip in the tree), we must divide the number of occurrences of $\mathcal{S}$ by the number of tips. A tip is also a  (very simple) shape $\mathcal{T}$ defined by the characteristic $\phi^\mathcal{T}(t)$ which equals 1 if $t$ is larger than the birth time of $x$. It follows that $E[\phi^\mathcal{T}(t)]=1$ for any $t>0$. If the tip characteristic is used in the denominator of equation (\ref{eq:characteristicRatio}), the denominator is $1/M$ and we find that the asymptotic frequency of $\mathcal{S}$ in a tree is
\begin{equation}\label{eq:shapeFrequency}
M\int_0^\infty\me^{-Mt}E\left[\phi^\mathcal{S}(t)\right]\text{d}t,
\end{equation}
with $\phi^\mathcal{S}:=\phi^\mathcal{S}_0$ to ensure consistency between the time variables.

\section{Cherries}
\subsection{Cherries in homogeneous processes}\label{sec:cherry-jccp}
\subsubsection*{Cherry characteristic}
At time $t$, a cherry $\mathcal{C}$ is the configuration made by an individual $x$ and its last descendant $(x,\xi_x(t))$ when the latter has no descendants, i.e. $\xi_{(x,\xi_x(t))}(t)=0$. It is called a cherry because it visually resembles a cherry (two tips of the tree joined at the same node). To be the ancestor of a cherry, $x$ must have at least one daughter (i.e. $\xi_x(t)\geq1$), and  $x$'s last daughter must have no descendants, i.e. $\xi_{(x,\xi_x(t))}(t)=0$. Accordingly, the cherry characteristic can be written as follows:
\begin{equation}\label{eq:cherry-characteristic}
	\phi^\mathcal{C}_x(t)=\begin{cases}1&\text{ if }\xi_x(t)\geq1\text{ and }\xi_{(x,\xi_x(t))}(t)=0 \\ 0&\text{ otherwise}\end{cases}.
\end{equation}
The total number of cherries in the tree is then: $Z^\mathcal{C}(t)=\sum_{x\in I}\phi^\mathcal{C}_x(t)$.
\subsubsection*{Derivation of $E\left[\phi^\mathcal{C}(t)\right]$}
In the homogeneous process, all individuals alive at a given time generate offspring at a constant rate $\beta$ and die at a constant rate $\delta$. In order to derive the asymptotic cherry frequency using equation (\ref{eq:shapeFrequency}) it is necessary to first derive $E\left[\phi^\mathcal{C}(t)\right]$. For this purpose it is convenient to use the JCCP because in the homogeneous setting the JCCP is the stochastic process that has almost everywhere derivative -1, which jumps at rate $\beta$ and whose jumps have random size exponentially distributed with parameter $\delta$ \cite{lambert2010contour}; in addition the process is sent back to the current time $t$ whenever it overshoots (reflection) and is killed upon hitting 0. Since a tree uniquely defines a JCCP and vice versa, then the law of the JCCP is also the law of the tree (modulo labelling of the tips).

\begin{figure}[t!]
	\begin{center}
		\includegraphics[width=\textwidth]{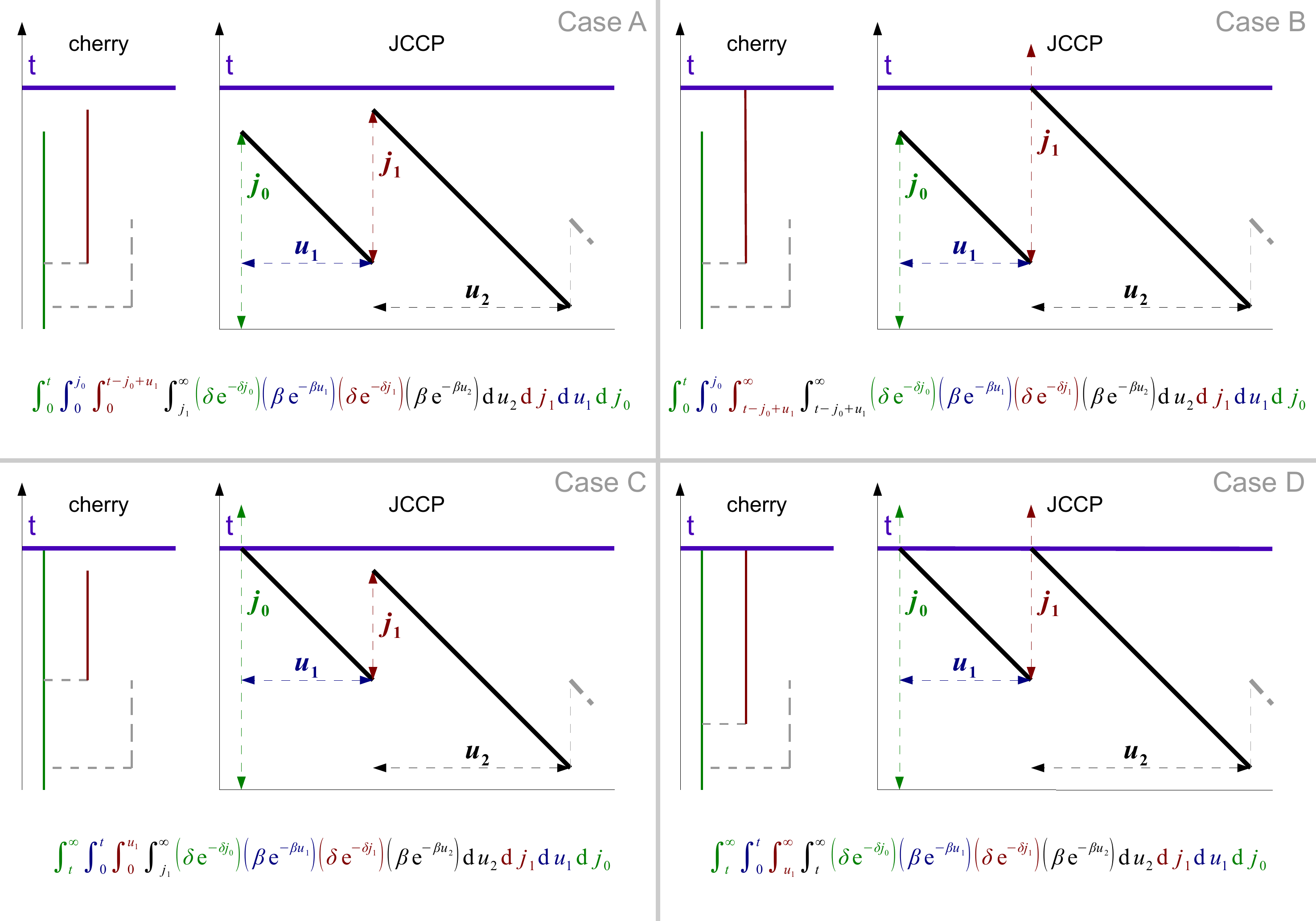}
	\end{center}
	\caption{Schematic for the calculation of $E[\phi^\mathcal{C}(t)]$. Because each of the two jumps $j_0$ and $j_1$ of the JCCP related to a cherry can overshoot above the current time $t$, four different cases have to be examined. The probability of each case can be derived with a 4-dimensional integral. In the computation, recall that the $j_i\sim Exp(\delta)$ and $u_i\sim Exp(\beta)$, for $i=0,1,2$.}\label{fig:cherry-jccp}
\end{figure}

In order to carry out analysis using the JCCP we use $u_i$ for the $i$-th inter-jump time and $j_{i-1}$ for the $i$-th jump size; the notation is depicted in Figures~\ref{fig:examplejccp} and \ref{fig:cherry-jccp}.
We aim to find a relationship among the $u$ and $j$ variables in the JCCP that is equivalent to the cherry characteristic (\ref{eq:cherry-characteristic}). Recall from section \ref{sec:shapeCharacteristic} that $\phi^\mathcal{C}(t)$ in Eq.~(\ref{eq:shapeFrequency}) is $\phi^\mathcal{C}_0(t)$. For this reason we will only focus on the part of the JCCP which corresponds to the ancestor and its last daughter: $j_0$, $u_1$, $j_1$, $u_2$.

First note that the ancestor has at least 1 daughter if and only if the JCCP does not hit 0 before the jump $j_1$. This ensures the presence of at least one daughter. Secondly, for the last daughter to have no descendants, the second low peak of the JCCP (the base of the third jump $j_2$) has to be lower than the first (the base of the second jump $j_1$).

Recall that the JCCP involves reflection, which means that when the JCCP overshoots, i.e. it jumps beyond the current time threshold $t$, the process is sent back to $t$. Because reflection  may happen on either of the two jumps, there are four distinct cases that give rise to a cherry (see Figure~\ref{fig:cherry-jccp}). Therefore, a cherry including the ancestor occurs only when the quartet $(j_0,u_1,j_1,u_2)$ is in one, and one only, of these four subsets of $\mathbb{R}_+^4$:
$$
A:\left\{\begin{aligned}
	&0<j_0\leq t \\
	&0<u_1<j_0\\
	&0<j_1\leq t-j_0+u_1\\
	&j_1< u_2<\infty\\
\end{aligned}\right., \hspace{0.1cm}
B:\left\{\begin{aligned}
	&0<j_0\leq t \\
	&0<u_1<j_0\\
	&t-j_0+u_1<j_1<\infty\\
	&t-j_0+u_1 < u_2<\infty\\
\end{aligned}\right., \hspace{0.1cm}
C:\left\{\begin{aligned}
	&t<j_0< \infty \\
	&0<u_1<t\\
	&0<j_1\leq u_1\\
	&j_1<u_2<\infty\\
\end{aligned}\right., \hspace{0.1cm}
D:\left\{\begin{aligned}
	&t<j_0< \infty \\
	&0<u_1<t\\
	&u_1<j_1<\infty\\
	&u_1<u_2<\infty\\
\end{aligned}\right..
 $$
 To find the probability that the ancestor mothers a cherry, we are left to measure the four sets above. Recall that $j_i\sim\text{Exp}(\delta)$ and $u_i\sim\text{Exp}(\beta)$, then $\text{d}(j_i)=\delta\me^{\delta j_i}\text{d}j_i$ and $\text{d}(u_i)=\beta\me^{-\beta u_i}\text{d}u_i$ for $i=0,1,2$. We derive the probability of each set:
\begin{displaymath}\begin{split}
A:\int_0^t\int_0^{j_0}\int_0^{t-j_0+u_1}\int_{j_1}^\infty \text{d}(u_2)\text{d}(j_1)\text{d}(u_1)\text{d}(j_0)=\frac{\beta\delta}{(\beta+\delta)^2}-\frac{2\delta}{2\beta+\delta}\me^{-\delta t}&+\frac{2\delta^2}{(\beta+\delta)^2}\me^{-(\beta+\delta)t}-\\
&-\frac{\beta\delta^2}{(2\beta+\delta)(\beta+\delta)^2}\me^{-2(\beta+\delta)t}
\end{split}\end{displaymath}

\begin{displaymath}\begin{split}
B:\int_0^t\int_0^{j_0}\int_{t-j_0+u_1}^{\infty}\int_{t-j_0+u_1}^{\infty}\text{d}(u_2)\text{d}(j_1)\text{d}(u_1)\text{d}(j_0)=\frac{\delta}{2\beta+\delta}\me^{-\delta t}-\frac{\delta}{\beta+\delta}\me^{-(\beta+\delta)t}+\frac{\beta\delta}{(2\beta+\delta)(\beta+\delta)}\me^{-2(\beta+\delta)t}
\end{split}\end{displaymath}

\begin{displaymath}\begin{split}
C:\int_t^\infty\int_0^{t}\int_0^{u_1}\int_{j_1}^\infty\text{d}(u_2)\text{d}(j_1)\text{d}(u_1)\text{d}(j_0)=\frac{\delta}{2\beta+\delta}\me^{-\delta t}-\frac{\delta}{\beta+\delta}\me^{-(\beta+\delta)t}+\frac{\beta\delta}{(\beta+\delta)(2\beta+\delta)}\me^{-2(\beta+\delta)t}
\end{split}\end{displaymath}

\begin{displaymath}\begin{split}
D:\int_t^\infty\int_0^{t}\int_{u_1}^\infty\int_{u_1}^\infty\text{d}(u_2)\text{d}(j_1)\text{d}(u_1)\text{d}(j_0) =\frac{\beta}{2\beta+\delta}\me^{-\delta t}-\frac{\beta}{2\beta+\delta}\me^{-2(\beta+\delta)t}
\end{split}\end{displaymath}
Now we only need to sum the four integrals:
 \begin{equation}\label{eq:ect}
E\left[\phi^\mathcal{C}(t)\right]=\frac{\beta\delta}{(\beta+\delta)^2}+\frac{\beta}{2\beta+\delta}\me^{-\delta t}-\frac{2\beta\delta}{(\beta+\delta)^2}\me^{-(\beta+\delta)t}-\frac{\beta^3}{(2\beta+\delta)(\beta+\delta)^2}\me^{-2(\beta+\delta)t}
\end{equation} 
Note that as $t\to\infty$, $E[c(t)]$ converges to $\p(\xi(\infty)\geq 1)\p(\xi(\infty)=0)$ which is the product of the probabilities of the two events that define the cherry characteristic. Moreover for $t=0$ the terms cancel and $E[c(t)]=0$; this confirms the impossibility to generate a cherry when no time has elapsed.

\subsubsection*{The cherries to tips ratio in homogeneous processes}
By ``cherries to tips ratio'' or CTR we indicate the limit of the frequency of the cherry shape $\mathcal{C}$ in a tree. Using equation (\ref{eq:shapeFrequency}):
\begin{equation}\label{eq:CTRintegral}
\frac{Z^\mathcal{C}(t)}{Z^\mathcal{T}(t)}\to CTR=M\int_0^\infty\me^{-Mt}E[\phi^\mathcal{C}(t)]\text{d}t.
\end{equation}
In homogeneous branching trees the Malthusian parameter is $M=\beta-\delta$ and it is positive because we assumed the process to be supercritical. The mean number of offspring for each individual, or basic reproduction number, is $R_0=\beta/\delta$ which is always greater than 1 under the supercritical assumption. Substituting equation (\ref{eq:ect}) we can compute the integral in equation (\ref{eq:CTRintegral}) to obtain $\frac{\beta}{(3\beta+\delta)(\beta-\delta)}$. Substitution of $M$ and $R_0$ gives
\begin{equation}\label{eq:ctt}
\text{CTR}=\frac{\beta}{3\beta+\delta}=\frac{R_0}{3R_0+1},
\end{equation}
Note that in the limit $\beta\to\infty$, that is the limit where the homogeneous tree tends to a Yule tree, the CTR tends to 1/3 which is a known result for Yule trees \cite{rosenberg2006mean,mckenzie2000distributions}. The methodology used to derive the result in equation (\ref{eq:ctt}) can be applied to different shapes and different choices of the branching process (i.e. non-homogeneous). When dealing with non-homogeneous trees we should bear in mind that the JCCP maintains its most important property (independence of $j_i$ and $u_i$) only when there is a constant birth rate. So for this approach to apply, the non-homogeneity must come from a non-exponential lifespan rather than a non-constant birth rate. Otherwise, evaluation of $E[\phi^\mathcal{S}]$ is likely to be challenging.

\subsection{Cherries in a non-homogeneous model}
Using the same approach as in section \ref{sec:cherry-jccp}, we found the cherry to tips ratio in a non-homogeneous model. If we choose the  life-span distribution to be a Gamma distribution with rate $\delta$ and shape 2, the Malthusian parameter and $R_0$ are given by
$$M=\frac{1}{2}\beta-\delta+\frac{1}{2}\sqrt{\beta^2+4\beta\delta}, ~~~~~ R_0=\frac{2\beta}{\delta}.$$
Using equation (\ref{eq:shapeFrequency}) after evaluating $E[\phi^\mathcal{C}(t)]$ in this model, the cherry to tips ratio is:
\begin{equation}
\begin{split}
CTR=&\biggr(512\Bigr(243R_0^{12}+243R_0^{23/2}\sqrt{R_0+8}+5103R_0^{11}+4131R_0^{21/2}\sqrt{R_0+8}+40851R_0^{10}+\\
&+26271R_0^{19/2}\sqrt{R_0+8}+160767R_0^9+80955R_0^{17/2}\sqrt{R_0+8}+338148R_0^8+\\
&+131184R_0^{15/2}\sqrt{R_0+8}+387448R_0^7+112912R_0^{13/2}\sqrt{R_0+8}+235072R_0^6+\\
&+49152R_0^{11/2}\sqrt{R_0+8}+68784R_0^5+9360R0^{9/2}\sqrt{R_0+8}+7616R_0^4+\\
&+512R_0^{7/2}\sqrt{R_0+8}+128R_0^3\Bigr)\biggr)\times\\
&\times\biggr(27R_0^5+27R_0^{9/2}\sqrt{R_0+8}+297R_0^4+189R_0^{7/2}\sqrt{R_0+8}+900R_0^3+360R_0^{5/2}\sqrt{R_0+8}+\\
&+968R_0^2+240R_0^{3/2}\sqrt{R_0+8}+368R_0+48\sqrt{R_0+8}\sqrt{R_0}+32\biggr)^{-1}\times\\
&\times\biggr(3R_0+\sqrt{R_0+8}\sqrt{R_0}\biggr)^{-2}\times\biggr(R_0+\sqrt{R_0+8} \sqrt{R_0}\biggr)^{-2}\times\biggr(5R_0+\sqrt{R_0+8}\sqrt{R_0}+4\biggr)^{-3}.
\end{split}
\end{equation}

\section{Pitchforks and double-cherries}
\subsubsection*{Pitchfork characteristics}
A pitchfork $\mathcal{P}$ is a configuration with three tips (illustrated in Figure~\ref{fig:summary}). A pitchfork is formed when an individual $x$' last  two daughters each have no descendants, or if her last daughter has only one descendant. The characteristic is written accordingly:
\begin{equation}\label{eq:pitchforkCharacteristic}
\phi^\mathcal{P}_x(t)=
\begin{cases}1&\text{ if }
	\left[\begin{array}{l}
	\xi_x(t)\geq2\text{ and }\xi_{(x,\xi_x(t))}(t)=0\text{ and }\xi_{(x,\xi_x(t)-1)}(t)=0 \\
	\text{ or } \\
	\xi_x(t)\geq1\text{ and }\xi_{(x,\xi_x(t))}(t)=1\text{ and }\xi_{(x,\xi_x(t),1)}(t)=0 
	\end{array}\right. \\
0&\text{otherwise}\end{cases}.
\end{equation}

\subsubsection*{The pitchfork to tips ratio in homogeneous processes}
In order to derive the pitchfork to tips ratio (PTR) we use equation (\ref{eq:shapeFrequency}). To evaluate $E[\phi^{\mathcal{P}}(t)]$ we use the JCCP as in section \ref{sec:cherry-jccp}: from the definition in equation (\ref{eq:pitchforkCharacteristic}) we determine the relationships among the first three peaks and three drops of the JCCP that correspond to a pitchfork in the tree. The resulting set in $\mathbb{R}^6$ is formed of 16 components that are measured with 6-dimensional integrals using the software \texttt{Maple} \cite{maple18}. There are two cases when a pitchfork is formed: the first extends the cherry and occurs when the ancestor has at least two daughters and the last two of them have no descendants. The second case occurs when the ancestor has at least one daughter and the ancestor's last daughters has one only daughter who does not have any further descendants. Because of possible overshooting, for each case there are 8 different sets that we should measure:

\begin{center}$
	\int_0^t
	\int_0^{j_0}
	\int_0^{t-j_0+u_1}
	\int_{j_1}^{j_0-u_1+j_1}
	\int_0^{t-j_0+u_1-j_1+u_2}
	\int_{j_2}^{\infty}\text{d}(u_3)\text{d}(j_2)\text{d}(u_2)\text{d}(j_1)\text{d}(u_1)\text{d}(j_0)
$\end{center}

\begin{center}$
	\int_0^{ t }
	\int_0^{j_0}
	\int_0^{ t-j_0+u_1}
	\int_{j_1}^{j_0-u_1+j_1}
	\int_{t-j_0+u_1-j_1+u_2}^{\infty}
	\int_{t-j_0+u_1-j_1+u_2}^{\infty}\text{d}(u_3)\text{d}(j_2)\text{d}(u_2)\text{d}(j_1)\text{d}(u_1)\text{d}(j_0)
$\end{center}

\begin{center}$
	\int_0^{ t }
	\int_0^{j_0}
	\int_{t-j_0+u_1}^{\infty}
	\int_{t-j_0+u_1}^{t}
	\int_0^{u_2}
	\int_{j_2}^{\infty}\text{d}(u_3)\text{d}(j_2)\text{d}(u_2)\text{d}(j_1)\text{d}(u_1)\text{d}(j_0)
$\end{center}

\begin{center}$
	\int_0^{ t }
	\int_0^{j_0}
	\int_{t-j_0+u_1}^{\infty}
	\int_{t-j_0+u_1}^{t}
	\int_{u_2}^{\infty}
	\int_{u_2}^{\infty}\text{d}(u_3)\text{d}(j_2)\text{d}(u_2)\text{d}(j_1)\text{d}(u_1)\text{d}(j_0)
$\end{center}

\begin{center}$
	\int_t^{ \infty }
	\int_0^{t}
	\int_0^{ u_1}
	\int_{j_1}^{t-u_1+j_1}
	\int_0^{u_1-j_1+u_2}
	\int_{j_2}^{\infty}\text{d}(u_3)\text{d}(j_2)\text{d}(u_2)\text{d}(j_1)\text{d}(u_1)\text{d}(j_0)
$\end{center}

\begin{center}$
	\int_t^{ \infty }
	\int_0^{t}
	\int_0^{ u_1}
	\int_{j_1}^{t-u_1+j_1}
	\int_{u_1-j_1+u_2}^{\infty}
	\int_{u_1-j_1+u_2}^{\infty}\text{d}(u_3)\text{d}(j_2)\text{d}(u_2)\text{d}(j_1)\text{d}(u_1)\text{d}(j_0)
$\end{center}

\begin{center}$
	\int_t^{ \infty }
	\int_0^{t}
	\int_{u_1}^{\infty}
	\int_{u_1}^{t}
	\int_0^{u_2}
	\int_{j_2}^{\infty}\text{d}(u_3)\text{d}(j_2)\text{d}(u_2)\text{d}(j_1)\text{d}(u_1)\text{d}(j_0)
$\end{center}

\begin{center}$
	\int_t^{ \infty }
	\int_0^{t}
	\int_{u_1}^{\infty}
	\int_{u_1}^{t}
	\int_{u_2}^{\infty}
	\int_{u_2}^{\infty}\text{d}(u_3)\text{d}(j_2)\text{d}(u_2)\text{d}(j_1)\text{d}(u_1)\text{d}(j_0)
$\end{center}

\begin{center}$
	\int_0^{ t }
	\int_0^{j_0}
	\int_0^{t-j_0+u_1}
	\int_0^{j_1}
	\int_0^{t-j_0+u_1-j_1+u_2}
	\int_{t-j_0+u_1}^{\infty}\text{d}(u_3)\text{d}(j_2)\text{d}(u_2)\text{d}(j_1)\text{d}(u_1)\text{d}(j_0)
$\end{center}

\begin{center}$
	\int_0^{ t }
	\int_0^{j_0}
	\int_0^{t-j_0+u_1}
	\int_0^{j_1}
	\int_{t-j_0+u_1-j_1+u_2}^{\infty}
	\int_{t-j_0+u_1}^{\infty}\text{d}(u_3)\text{d}(j_2)\text{d}(u_2)\text{d}(j_1)\text{d}(u_1)\text{d}(j_0)
$\end{center}

\begin{center}$
	\int_0^{t }
	\int_0^{j_0}
	\int_{t-j_0+u_1}^{\infty}
	\int_0^{t-j_0+u_1}
	\int_0^{u_2}
	\int_{t-u_2+j_2-j_0+u_1}^{\infty}\text{d}(u_3)\text{d}(j_2)\text{d}(u_2)\text{d}(j_1)\text{d}(u_1)\text{d}(j_0)
$\end{center}

\begin{center}$
	\int_0^{t }
	\int_0^{j_0}
	\int_{t-j_0+u_1}^{\infty}
	\int_0^{t-j_0+u_1}
	\int_{u_2}^{\infty}
	\int_{t-j_0+u_1}^{\infty}\text{d}(u_3)\text{d}(j_2)\text{d}(u_2)\text{d}(j_1)\text{d}(u_1)\text{d}(j_0)
$\end{center}

\begin{center}$
	\int_t^{\infty}
	\int_0^{t}
	\int_0^{u_1}
	\int_0^{j_1}
	\int_0^{u_1-j_1+u_2}
	\int_{j_1-u_2+j_2}^{\infty}\text{d}(u_3)\text{d}(j_2)\text{d}(u_2)\text{d}(j_1)\text{d}(u_1)\text{d}(j_0)
$\end{center}

\begin{center}$
	\int_t^{\infty}
	\int_0^{t}
	\int_0^{u_1}
	\int_0^{j_1}
	\int_{u_1-j_1+u_2}^{\infty}
	\int_{u_1}^{\infty}\text{d}(u_3)\text{d}(j_2)\text{d}(u_2)\text{d}(j_1)\text{d}(u_1)\text{d}(j_0)
$\end{center}

\begin{center}$
	\int_t^{\infty}
	\int_0^{t}
	\int_{u_1}^{\infty}
	\int_0^{u_1}
	\int_0^{u_2}
	\int_{u_1-u_2+j_2}^{\infty}\text{d}(u_3)\text{d}(j_2)\text{d}(u_2)\text{d}(j_1)\text{d}(u_1)\text{d}(j_0)
$\end{center}

\begin{center}$
	\int_t^{\infty}
	\int_0^{t}
	\int_{u_1}^{\infty}
	\int_0^{u_1}
	\int_{u_2}^{\infty}
	\int_{u_1}^{\infty}\text{d}(u_3)\text{d}(j_2)\text{d}(u_2)\text{d}(j_1)\text{d}(u_1)\text{d}(j_0)
$\end{center}
To evaluate the integrals, recall that $j_i\sim\text{Exp}(\delta)$ and $u_i\sim\text{Exp}(\beta)$, then $\text{d}(j_i)=\delta\me^{\delta j_i}\text{d}j_i$ and $\text{d}(u_i)=\beta\me^{-\beta u_i}\text{d}u_i$ for $i=0,1,2,3$. The expression for the pitchfork to tips ratio in the homogeneous model is:
\begin{align}\label{eq:PTR}
\nonumber PTR&=\frac{3(\beta-\delta)(\beta+\delta)\beta^2}{(2\beta+\delta)(\beta-\delta)(3\beta+\delta)^2}\\
&=\frac{3(R_0+1)(R_0^2)}{(2R_0+1)(3R_0+1)^2}.
\end{align}
Note that the limit of $PTR$ as $R_0\to\infty$, where the homogeneous model tends to the Yule model, is $1/6$, which corresponds to Rosenberg's result for the Yule model in \cite{rosenberg2006mean}.

\subsection{Frequency of the symmetric configuration with 4 tips in homogeneous processes}
\subsubsection*{The characteristic}
Let's consider a configuration $\mathcal{S}$ of four tips organized in two cherries. In order to count such configurations we define a characteristic which assigns 1 to every individual $x$ which is both the ancestor and a tip of the configuration. This happens if among the daughters (at least 2) of $x$, the last has no descendants and the second last has one only descendant. Equivalently, the ancestor mothers a cherry and so does the ancestor's second last daughter.
\begin{equation}\label{eq:DCCharacteristic}
\phi^\mathcal{S}_x(t)=
\begin{cases}
1&\text{ if }\xi_x(t)\geq2\text{ and }\xi_{(x,\xi_x(t))}(t)=0\text{ and }\xi_{(x,\xi_x(t)-1)}(t)=1\text{ and }\xi_{(x,\xi_x(t)-1,1)}=0\\
0&\text{ o.w.}\end{cases}.
\end{equation}
\subsubsection*{Derivation of the asymptotic frequency}
We use equation (\ref{eq:shapeFrequency}) to evaluate the asymptotic frequency of $\mathcal{S}$. As in the previous sections we evaluate $E[\phi^\mathcal{S}(t)]$ using the JCCP. Its derivation involves 16 8-dimensional integrals:
\begin{center}
\resizebox{\textwidth}{!}{$
	\int_0^{t}
	\int_0^{j_0}
	\int_0^{ t-j_0+u_1}
	\int_{j_1}^{j_0-u_1+j_1}
	\int_0^{t-j_0+u_1-j_1+u_2}
	\int_0^{j_2}
	\int_0^{t-j_0+u_1-j_1+u_2-j_2+u_3}
	\int_{j_3-u_3+j_2}^{\infty}
	\text{d}(u_4)\text{d}(j_3)\text{d}(u_3)\text{d}(j_2)\text{d}(u_2)\text{d}(j_1)\text{d}(u_1)\text{d}(j_0)
$}
\end{center}

\begin{center}
\resizebox{\textwidth}{!}{$
	\int_0^{ t }
	\int_0^{j_0}
	\int_0^{ t-j_0+u_1}
	\int_{j_1}^{j_0-u_1+j_1}
	\int_0^{t-j_0+u_1-j_1+u_2}
	\int_0^{j_2}
	\int_{t-j_0+u_1-j_1+u_2-j_2+u_3}^{\infty}
	\int_{t-j_0+u_1-j_1+u_2}^{\infty}
	\text{d}(u_4)\text{d}(j_3)\text{d}(u_3)\text{d}(j_2)\text{d}(u_2)\text{d}(j_1)\text{d}(u_1)\text{d}(j_0)
$}
\end{center}

\begin{center}
\resizebox{\textwidth}{!}{$
	\int_0^{ t }
	\int_0^{j_0}
	\int_0^{ t-j_0+u_1}
	\int_{j_1}^{j_0-u_1+j_1}
	\int_{t-j_0+u_1-j_1+u_2}^{\infty}
	\int_0^{t-j_0+u_1-j_1+u_2}
	\int_0^{u_3}
	\int_{t-u_3+j_3-j_0+u_1-j_1+u_2}^{\infty}
	\text{d}(u_4)\text{d}(j_3)\text{d}(u_3)\text{d}(j_2)\text{d}(u_2)\text{d}(j_1)\text{d}(u_1)\text{d}(j_0)
$}
\end{center}

\begin{center}
\resizebox{\textwidth}{!}{$
	\int_0^{ t }
	\int_0^{j_0}
	\int_0^{ t-j_0+u_1}
	\int_{j_1}^{j_0-u_1+j_1}
	\int_{t-j_0+u_1-j_1+u_2}^{\infty}
	\int_0^{t-j_0+u_1-j_1+u_2}
	\int_{u_3}^{\infty}
	\int_{t-j_0+u_1-j_1+u_2}^{\infty}
	\text{d}(u_4)\text{d}(j_3)\text{d}(u_3)\text{d}(j_2)\text{d}(u_2)\text{d}(j_1)\text{d}(u_1)\text{d}(j_0)
$}
\end{center}

\begin{center}
{$
	\int_0^{ t }
	\int_0^{j_0}
	\int_{t-j_0+u_1}^{\infty}
	\int_{t-j_0+u_1}^{t}
	\int_0^{u_2}
	\int_0^{j_2}
	\int_0^{u_2-j_2+u_3}
	\int_{j_2-u_3+j_3}^{\infty}
	\text{d}(u_4)\text{d}(j_3)\text{d}(u_3)\text{d}(j_2)\text{d}(u_2)\text{d}(j_1)\text{d}(u_1)\text{d}(j_0)
$}
\end{center}

\begin{center}
{$
	\int_0^{ t }
	\int_0^{j_0}
	\int_{t-j_0+u_1}^{\infty}
	\int_{t-j_0+u_1}^{t}
	\int_0^{u_2}
	\int_0^{j_2}
	\int_{u_2-j_2+u_3}^{\infty}
	\int_{u_2}^{\infty}
	\text{d}(u_4)\text{d}(j_3)\text{d}(u_3)\text{d}(j_2)\text{d}(u_2)\text{d}(j_1)\text{d}(u_1)\text{d}(j_0)
$}
\end{center}

\begin{center}
{$
	\int_0^{ t }
	\int_0^{j_0}
	\int_{t-j_0+u_1}^{\infty}
	\int_{t-j_0+u_1}^{t}
	\int_{u_2}^{\infty}
	\int_0^{u_2}
	\int_0^{u_3}
	\int_{u_2-u_3+j_3}^{\infty}
	\text{d}(u_4)\text{d}(j_3)\text{d}(u_3)\text{d}(j_2)\text{d}(u_2)\text{d}(j_1)\text{d}(u_1)\text{d}(j_0)
$}
\end{center}

\begin{center}
{$
	\int_0^{ t }
	\int_0^{j_0}
	\int_{t-j_0+u_1}^{\infty}
	\int_{t-j_0+u_1}^{t}
	\int_{u_2}^{\infty}
	\int_0^{u_2}
	\int_{u_3}^{\infty}
	\int_{u_2}^{\infty}
	\text{d}(u_4)\text{d}(j_3)\text{d}(u_3)\text{d}(j_2)\text{d}(u_2)\text{d}(j_1)\text{d}(u_1)\text{d}(j_0)
$}
\end{center}

\begin{center}
{$
	\int_t^{ \infty }
	\int_0^{t}
	\int_0^{ u_1}
	\int_{j_1}^{t-u_1+j_1}
	\int_0^{u_1-j_1+u_2}
	\int_0^{j_2}
	\int_0^{u_1-j_1+u_2-j_2+u_3}
	\int_{j_2-u_3+j_3}^{\infty}
	\text{d}(u_4)\text{d}(j_3)\text{d}(u_3)\text{d}(j_2)\text{d}(u_2)\text{d}(j_1)\text{d}(u_1)\text{d}(j_0)
$}
\end{center}

\begin{center}
{$
	\int_t^{ \infty }
	\int_0^{t}
	\int_0^{ u_1}
	\int_{j_1}^{t-u_1+j_1}
	\int_0^{u_1-j_1+u_2}
	\int_0^{j_2}
	\int_{u_1-j_1+u_2-j_2+u_3}^{\infty}
	\int_{u_1-j_1+u_2}^{\infty}
	\text{d}(u_4)\text{d}(j_3)\text{d}(u_3)\text{d}(j_2)\text{d}(u_2)\text{d}(j_1)\text{d}(u_1)\text{d}(j_0)
$}
\end{center}

\begin{center}
{$
	\int_t^{ \infty }
	\int_0^{t}
	\int_0^{ u_1}
	\int_{j_1}^{t-u_1+j_1}
	\int_{u_1-j_1+u_2}^{\infty}
	\int_0^{u_1-j_1+u_2}
	\int_0^{u_3}
	\int_{u_1-j_1+u_2-u_3+j_3}^{\infty}
	\text{d}(u_4)\text{d}(j_3)\text{d}(u_3)\text{d}(j_2)\text{d}(u_2)\text{d}(j_1)\text{d}(u_1)\text{d}(j_0)
$}
\end{center}

\begin{center}
{$
	\int_t^{ \infty }
	\int_0^{t}
	\int_0^{ u_1}
	\int_{j_1}^{t-u_1+j_1}
	\int_{u_1-j_1+u_2}^{\infty}
	\int_0^{u_1-j_1+u_2}
	\int_{u_3}^{\infty}
	\int_{u_1-j_1+u_2}^{\infty}
	\text{d}(u_4)\text{d}(j_3)\text{d}(u_3)\text{d}(j_2)\text{d}(u_2)\text{d}(j_1)\text{d}(u_1)\text{d}(j_0)
$}
\end{center}

\begin{center}
{$
	\int_t^{ \infty }
	\int_0^{t}
	\int_{u_1}^{\infty}
	\int_{u_1}^{t}
	\int_0^{u_2}
	\int_0^{j_2}
	\int_0^{u_2-j_2+u_3}
	\int_{j_2-u_3+j_3}^{\infty}
	\text{d}(u_4)\text{d}(j_3)\text{d}(u_3)\text{d}(j_2)\text{d}(u_2)\text{d}(j_1)\text{d}(u_1)\text{d}(j_0)
$}
\end{center}

\begin{center}
{$
	\int_t^{ \infty }
	\int_0^{t}
	\int_{u_1}^{\infty}
	\int_{u_1}^{t}
	\int_0^{u_2}
	\int_0^{j_2}
	\int_{u_2-j_2+u_3}^{\infty}
	\int_{u_2}^{\infty}
	\text{d}(u_4)\text{d}(j_3)\text{d}(u_3)\text{d}(j_2)\text{d}(u_2)\text{d}(j_1)\text{d}(u_1)\text{d}(j_0)
$}
\end{center}

\begin{center}
{$
	\int_t^{ \infty }
	\int_0^{t}
	\int_{u_1}^{\infty}
	\int_{u_1}^{t}
	\int_{u_2}^{\infty}
	\int_0^{u_2}
	\int_0^{u_3}
	\int_{t-u_3+j_3-u_2}^{\infty}
	\text{d}(u_4)\text{d}(j_3)\text{d}(u_3)\text{d}(j_2)\text{d}(u_2)\text{d}(j_1)\text{d}(u_1)\text{d}(j_0)
$}
\end{center}

\begin{center}
{$
	\int_t^{ \infty }
	\int_0^{t}
	\int_{u_1}^{\infty}
	\int_{u_1}^{t}
	\int_{u_2}^{\infty}
	\int_0^{u_2}
	\int_{u_3}^{\infty}
	\int_{u_2}^{\infty}
	\text{d}(u_4)\text{d}(j_3)\text{d}(u_3)\text{d}(j_2)\text{d}(u_2)\text{d}(j_1)\text{d}(u_1)\text{d}(j_0)
$}
\end{center}
To evaluate the integrals, recall that $j_i\sim\text{Exp}(\delta)$ and $u_i\sim\text{Exp}(\beta)$, then $\text{d}(j_i)=\delta\me^{\delta j_i}\text{d}j_i$ and $\text{d}(u_i)=\beta\me^{-\beta u_i}\text{d}u_i$ for $i=0,1,2,3,4$. The asymptotic frequency of $\mathcal{S}$ is given by:
\begin{equation}\label{eq:DCTR}
\begin{split}&\frac{1}{4}(2592R_0^9+11556R_0^8+18279R_0^7+13899R_0^6+4799R_0^5-65R_0^4-546R_0^3-114R_0^2)\\
&(19440R_0^9+91044R_0^8+187488R_0^7+222741R_0^6+168180R_0^5+83666R_0^4+27416R_0^3+\\
&+5705R_0^2+684R_0+36)^{-1}
\end{split}
\end{equation}

\section{Conclusion}
\begin{figure}[t]
	\begin{center}
		\includegraphics[width=\textwidth]{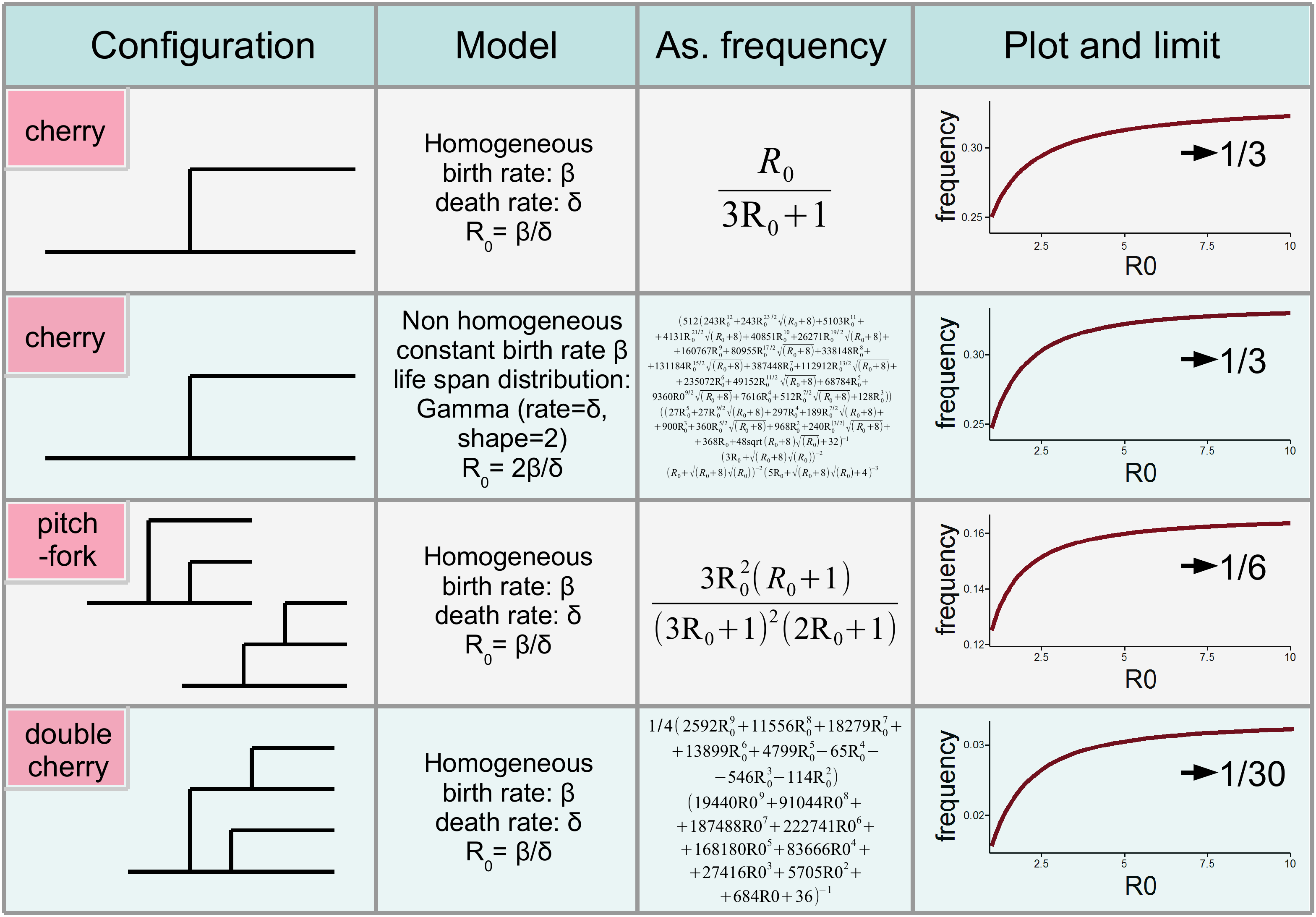}
	\end{center}
	\caption{Summary of the analytical results.}\label{fig:summary}
\end{figure}

In this paper we have presented a novel technique to compute the frequency of any shape configuration in a tree. These frequencies, namely the ratio of the number of occurrences of a specific shape configuration to the number of tips in the treee, converge in probability to the expression in Eq.~(\ref{eq:shapeFrequency}) as $t\rightarrow \infty$. We have applied the technique to evaluate the asymptotic frequency of cherries in the homogeneous process and in a non-homogeneous process (with non-exponential death/recovery rates). For the homogeneous tree, we have also derived the frequency of pitchforks and of the symmetric shape with four tips (double cherry). In Figure~\ref{fig:summary} we present a summary of the results. 

\newpage
\bibliographystyle{apt}
\bibliography{../../../../mybib/mybib}

\end{document}